\documentclass[aip,preprint]{revtex4-2}
\usepackage{tikz}
\usepackage{pgfplots}
\usepackage{pgfplotstable}
\pgfplotsset{width=10cm,compat=1.9}
\usepackage[cp1251]{inputenc}
\usepackage[T2A]{fontenc}
\usepackage[english]{babel}
\usepackage{amssymb,latexsym,amsmath,amscd}
\usepackage{graphicx,color,framed}
\usepackage{mathrsfs}
\usepackage{bm}
\usepackage{appendix}
\usepackage{tabu}
\usepackage{tikz}
\usepackage{xcolor}
\usepackage{siunitx}
\usepackage{empheq}
\usepackage{float}
\usepackage{subcaption}
\usepackage{xr}
\usepackage{siunitx}
\usepackage{microtype}
\usepackage{amsfonts}
\usepackage{hyperref}
\usepackage{gensymb}
\usepackage[normalem]{ulem}
\graphicspath{{figures/}}

\makeatletter
\newcommand*{\addFileDependency}[1]{
  \typeout{(#1)}
  \@addtofilelist{#1}
  \IfFileExists{#1}{}{\typeout{No file #1.}}
}
\makeatother

\begin{document}

\author{\firstname{Yury A.} \surname{Budkov}}
\email[]{ybudkov@hse.ru}
\affiliation{Laboratory of Computational Physics, HSE University, Tallinskaya st. 34, 123458 Moscow, Russia}
\affiliation{School of Applied Mathematics, HSE University, Tallinskaya st. 34, 123458 Moscow, Russia}
\affiliation{Laboratory of Multiscale Modeling of Molecular Systems, G.A. Krestov Institute of Solution Chemistry of the Russian Academy of Sciences, 153045, Akademicheskaya st. 1, Ivanovo, Russia}
\author{\firstname{Petr E.} \surname{Brandyshev}}
\affiliation{Laboratory of Computational Physics, HSE University, Tallinskaya st. 34, 123458 Moscow, Russia}
\affiliation{School of Applied Mathematics, HSE University, Tallinskaya st. 34, 123458 Moscow, Russia}

\title{First-principle theory of the Casimir screening effect~\footnote{The authors dedicate this article to the memory of Professor Rudolf Podgornik, who passed away suddenly, leaving a significant void in the field of theoretical physics. His contributions and discussions have enriched not only the current work, but also continue to inspire our own efforts.}}
\begin{abstract}
In this paper, we use the formalism of finite-temperature quantum field theory to investigate the Casimir force between flat, ideally conductive surfaces containing confined, but mobile ions. We demonstrate that, in the Gaussian approximation, the ionic fluctuations contribute separately from the electromagnetic fluctuations that are responsible for the standard Casimir effect. This is in line with the "separation hypothesis", has been previously applied on a purely intuitive basis. Our analysis demonstrates the significance of calculating the zero Matsubara frequency component in the electromagnetic contribution, using the formula developed by Schwinger et al., as opposed to the approach based on Lifshitz theory used by other researchers. 
\end{abstract}

\maketitle
\section{Introduction}

The Casimir effect, named after Hendrik Casimir, who first predicted it in 1948~\cite{casimir1948attraction}, is a phenomenon that occurs when two neutral metal surfaces are brought very close together in vacuum at zero temperature. The extension of Casimir's result to arbitrary temperatures and general dielectric media was accomplished by Lifshitz et al. ~\cite{lifshitz1992theory,dzyaloshinskii1961general}, with subsequent refinements by Schwinger et al.~\cite{schwinger1978casimir} and others~\cite{barash1975electromagnetic,mahanty1976dispersion,parsegian2005van,podgornik2003reformulation,bordag2001new,klimchitskaya2023casimir}.  Casimir effect is caused by the quantum fluctuations of electromagnetic fields in vacuum that generate a difference in the ground state energy density between the inside and outside of the metal surfaces, creating a net attraction between them. Although the Casimir force is very weak, it has been detected and experimentally quantified, and has been shown to play a crucial role in nanotechnology, particularly in the development of microelectromechanical systems (MEMS) and other nano devices~\cite{chan2001quantum,palasantzas2020applications,elsaka2024casimir}. Overall, the Casimir effect is a prime example of the complex and unintuitive nature of quantum physics~\cite{barash1975electromagnetic,bordag2001new}. It demonstrates how electromagnetic field fluctuations permeate empty space, generating long-range interactions between bodies that have a significant impact on many everyday phenomena.

Despite the significant progress made in the theoretical understanding of the Casimir effect~\cite{barash1975electromagnetic,mahanty1976dispersion,parsegian2005van,bordag2001new}, there are still some unresolved issues that need to be addressed.

One of the unresolved problems in the theory of the Casimir effect is the Casimir force in the presence of mobile charged particles (ions) between flat perfectly conductive walls. In early theoretical studies~\cite{barouch1973interdependence, gorelkin1973calculation, barnes1975statistical, ninham1997ion} it was assumed that electromagnetic field fluctuations responsible for the attractive Casimir force could be treated separately from the pure image effects of the metallic slabs, which influence the behavior of the electrolyte ions. This assumption was referred to as the "separation hypothesis" ~\cite{jancovici2004screening}. Following the separation hypothesis, two long-range forces cancel each other out exactly~\cite{gorelkin1973calculation, barnes1975statistical, ninham1997ion}. This canceling could be referred to as the "Casimir screening effect"~\cite{jancovici2005casimir}. Despite the plausibility of this hypothesis, it has not been confirmed by the fundamental principles of quantum electrodynamics.

We would like to emphasize that the screening of the Casimir force caused by the presence of ions in the space between metal walls is different from the screening related to the presence of mass within the quanta of the vector field within the framework of the Proca model~\cite{barton1984casimir,barton1985casimir,teo2010casimir}. Unlike 'ionic screening', 'mass screening', which is essentially a quantum effect, occurs for all terms in the Casimir force expansion in the Matsubara modes~\cite{barton1984casimir}.

The second issue is related to the behavior of the Casimir force at high temperatures. When the dielectric function of the walls material approaches infinity ($\epsilon(\omega) \to \infty$), the zero-Matsubara frequency term in the Lifshitz formula is not uniquely defined. Its value depends on the order in which the limits of the dielectric function and the Matsubara frequency (as $\omega_n\to 0$) are taken. To obtain the ideal-conductor result based on electrostatic boundary conditions, Schwinger et al. postulated the following order~\cite{schwinger1978casimir}: $\epsilon(\omega) \to \infty$ and then $\omega_n\to 0$. Thus, they obtained twice as high value of the high temperature limit of the Casimir force than what was obtained by Lifshitz et al.~\cite{dzyaloshinskii1961general}.  Jancovici and $\text{\v{S}}$amaj~\cite{jancovici2004screening,jancovici2005casimir} and independently Buenzli and Martin ~\cite{buenzli2005microscopic} questioned the result of Schwinger et al., modeling the Casimir interaction of metal walls as the interaction of half-spaces containing equilibrium plasma comprised of classical ions. Using methods based on the formalism of correlation functions~\cite{buenzli2005microscopic} and the theory of linear response~\cite{jancovici2004screening}, the Casimir force was obtained, found to be half the value reported by Schwinger et al., in accordance with the Lifshitz limiting regime. Later, Buenzli and Martin~\cite{buenzli2008microscopic} reproduced their result assuming that metals consist of mobile quantum charges in thermal equilibrium with the photon field at a fixed temperature. Recently, Brandyshev and Budkov~\cite{brandyshev2024finite} following the finite-temperature quantum field theory~\cite{kapusta2007finite}, directly reproduced Schwinger's result for the Casimir force. It is important to note that their first-principle quantum electrodynamics-based approach does not require the assumption of the order of the aforementioned limits, which may indicate the correctness of the result obtained by Schwinger and coworkers. Below we provide more detailed analysis.

The aim of this study is to conduct an analysis of these issues, considering their interconnections within the finite-temperature quantum field theory approach.

\section{Theory}

We assume that there are $N_{+}$ positively charged point-like ions (cations) and $N_{-}$ negatively charged point-like ions (anions) in a slit-like pore (Coulomb gas). The ions have charges of $q_{+}$ and $q_{-}$, respectively, and the walls of the pore are ideally conductive and electrically neutral.  The distance between the walls is $L$. We assume that the system, which consists of ions, conductive walls, and a quantum electromagnetic field, is in thermodynamic equilibrium at a specific temperature $T$. The total ionic charge is zero, i.e. $q_{+}N_{+}+q_{-}N_{-}=0$, and  the system is electroneutral.

Thus, we start from the partition function of the described system with a fixed gauge (see Appendix A)
\begin{equation}
\label{Z}
Z=\oint \mathcal{D}A|D|\oint d\sigma[\{\bold{r}_{i}^{(+)}\}]\oint d\sigma[\{\bold{r}_{j}^{(-)}\}]\exp\left[S_{EM}[A]-i\int\limits_{0}^{\beta}d\tau\int d^3\mathbf{x} (\hat{\rho}\varphi-\hat{\bold{j}}\cdot\bold{A})\right],
\end{equation}
where $A_{\mu}=(-\varphi,\bold{A})$ is the 4-potential of the electromagnetic field with effective action
\begin{equation}
S_{EM}[A]=\int\limits_{0}^{\beta}d\tau\int d^3\mathbf{x} \; \mathscr{L}_{\text{eff}},
\end{equation}
\begin{equation}\label{}
\begin{aligned}
\mathscr{L}_{\text{eff}} = \frac{1}{2}A_{\mu}\square A_{\mu} + \frac{1-\alpha}{2}\bigg(\partial_{\mu}A_{\mu}\bigg)^{2},
\end{aligned}
\end{equation}
$|D|=\text{Det}\left(-\square\right)$ is the functional determinant and $\alpha$ is the gauge parameter determining the family of similar gauges~\cite{weinberg1995quantum}; we have also introduced the integration measures over the paths, $\bold{r}_{i}^{(a)}(\tau)$, of ions in the imaginary 'time' $\tau$, i.e.
\begin{equation}
\oint d\sigma[\{\bold{r}_{i}^{(a)}\}](\dots) = \oint \prod\limits_{i=1}^{N_{a}}\frac{\mathcal{D}\bold{r}_{i}^{(a)}}{C_{a}}\exp\left[-\int\limits_{0}^{\beta}d\tau\frac{m_{a}(\dot{\bold{r}}_{i}^{(a)}(\tau))^2}{2}\right](\dots), ~a=\pm,
\end{equation}
where $m_a$ is the mass of ion $a$,  and $C_{a}=\oint \mathcal{D}\bold{r}_{i}^{(a)} e^{-\int\limits_{0}^{\beta}d\tau\frac{m_{a}\left(\dot{\bold{r}}_{i}^{(a)}(\tau)\right)^2}{2}}$ are the normalization constants; the symbol $\oint$ means that functional integration takes into account periodic boundary conditions for the integration variables,
\begin{equation}\label{1.29}
A_{\mu}(0,\mathbf{x})=A_{\mu}(\beta, \mathbf{x}), \quad \mu=0,1,2,3,
\end{equation}
\begin{equation}\label{1.29}
\bold{r}_{i}^{(a)}(0)=\bold{r}_{i}^{(a)}(\beta), \quad \beta=\frac{1}{T}.
\end{equation}

We would like to point out that the partition function, which is the trace of the density matrix, can be formally obtained from the quantum amplitude of a system written in terms of a functional integral, by performing a Wick rotation from time to inverse temperature, $\tau \to -i\tau$~\cite{feynman2010quantum,zinn2010path}. Note also that we use the Euclidean metric for D'Alembert operator, i.e. $\square = \partial_{\tau}^{2} + \Delta$ and a unit system, where $k_{B}=\hbar=c=1$ ($k_B$ is the Boltzmann constant, $\hbar$ is the Planck constant, and $c$ is the speed of light).

The microscopic charge density and current density of ions can be determined as follows
\begin{equation}
\label{j0}
\hat{\rho}(\tau,\bold{x})=\sum\limits_{a=\pm}q_{a}\sum\limits_{i=1}^{N_{a}}\delta\left(\bold{x}-\bold{r}_{i}^{(a)}(\tau)\right),
\end{equation}
\begin{equation}
\label{j}
\hat{\bold{j}}(\tau,\bold{x})=\sum\limits_{a=\pm}q_{a}\sum\limits_{j=1}^{N_{a}}\dot{\bold{r}}_{i}^{(a)}(\tau)\delta\left(\bold{x}-\bold{r}_{j}^{(a)}(\tau)\right).
\end{equation}
It is easy to check that the electric charge and current density obey the conservation law, $\partial_{\tau}\hat{\rho}+\nabla\cdot \hat{\bold{j}}=0$. Taking into account (\ref{j0}) and (\ref{j}), we can rewrite the partition function (\ref{Z}) in the form
\begin{equation}
\label{Z_tot}
Z=\oint \mathcal{D}A   |D|\exp\left[S_{EM}[A]\right] \prod\limits_{a=\pm}\left[\oint \frac{\mathcal{D}\bold{r}_a}{C_{a}}e^{-\int\limits_{0}^{\beta}d\tau\left(\frac{m_{a}\dot{\bold{r}}_a^2(\tau)}{2}+iq_a \varphi(\tau,\bold{r}_a(\tau))-iq_a \dot{\bold{r}}_a(\tau)\cdot\bold{A}(\tau,\bold{r}_a(\tau))\right)}\right]^{N_a}.
\end{equation}
Now, let us rewrite the 'paths' of ions as $\bold{r}_a(\tau) = \bold{x}_a+\boldsymbol{\xi}_a(\tau)$, where $\bold{x}_a=\beta^{-1}\int\limits_{0}^{\beta}d\tau\, \bold{r}_a(\tau)$ is the average position of particle, and variables $\boldsymbol{\xi}_{a}(\tau)$ describe its quantum 'smoothing'. Considering the ions as classical particles, that is, neglecting the terms containing the variables $\boldsymbol{\xi}_a(\tau)$ and $\dot{\boldsymbol{\xi}}_a(\tau)$, we can obtain
\begin{equation}
\label{Z_cl}
Z=\oint \mathcal{D}A |D| \exp\left[S_{EM}[A]\right] \prod\limits_{a=\pm}\left[\int\frac{d\bold{x}}{V}e^{-i\int\limits_{0}^{\beta}d\tau q_a \varphi(\tau,\bold{x})}\right]^{N_a},
\end{equation}
where $V$ is the volume of system and we take into account that $C_a=V\int \mathcal{D}\boldsymbol{\xi}\, e^{-\int\limits_{0}^{\beta}d\tau\frac{m_{a}\dot{\boldsymbol{\xi}}^2(\tau)}{2}}$. The ions can be treated as classical particles if their thermal wavelengths, $\lambda_a$, are much smaller than the average inter-ionic distances, which is always the case \cite{feynman2018statistical}.

In the thermodynamic limit $N_a\to \infty$, $V\to \infty$, $N_a/V\to n_a$, where $n_a$ are the average ionic concentrations, we can obtain~\cite{budkov2015modified,budkov2020statistical}
\begin{equation}
\label{Z_cl_2}
Z=\oint \mathcal{D}\varphi\mathcal{D}\bold{A}|D| \exp\left[S_{EM}[\varphi,\bold{A}]+S_{ion}[\varphi]\right],
\end{equation}
where we have used that $\mathcal{D}A=\mathcal{D}\varphi\mathcal{D}\bold{A}$ and introduced the contribution of the ions to the effective action
\begin{equation}
\label{Sion}
S_{ion}[\varphi]=\sum\limits_{a=\pm}n_a\int d\bold{x}\left(e^{-i\int\limits_{0}^{\beta}d\tau q_a \varphi(\tau,\bold{x})}-1\right).
\end{equation}
As seen from eqs. (\ref{Z_cl_2}) and (\ref{Sion}), for pure classical ions, the vector potential drops out from the ionic part of effective action. In other words, classical charged particles in equilibrium can only affect fluctuations in the electric potential $\varphi$, which is, of course, in accordance with the Bohr-Van Leeuwen theorem~\cite{savoie2015rigorous,kaufman2024lectures}.

To proceed with the calculation of the partition function (\ref{Z_cl_2}), we need to specify a gauge. The simplest one is the Feynman gauge, for which $\alpha=1$. In this case, the kinetic terms for all the fields $A_\mu$ are in canonical form, meaning they are orthogonalized and uncoupled. Let us calculate the partition function (\ref{Z_cl_2}) in the Gaussian approximation. Expanding the ionic part of the effective action ($S_{ion}$) in a power series of $\varphi$ and neglecting terms higher than second order, we get
\begin{equation}
\label{Z_cl_3}
Z=\oint \mathcal{D}\varphi\mathcal{D}\bold{A}|D| e^{\frac{1}{2}\int\limits_{0}^{\beta}d\tau\int d^3\mathbf{x}A_{\mu}\square A_{\mu}-\frac{1}{2}\sum\limits_{a=\pm}q_a^2n_{a}\int\limits_{0}^{\beta}d\tau\int\limits_{0}^{\beta}d\tau'\int d\bold{x}\varphi(\tau,\bold{x})\varphi(\tau',\bold{x})}.
\end{equation}
By calculating the Gaussian functional integral, taking into account the boundary conditions for an ideal metal (see Appendix B and ref. \cite{brandyshev2024finite}), we can obtain the surface free energy density
\begin{equation}
 f=-\frac{1}{\beta \mathcal{A}}\ln {Z}, 
\end{equation}
where $\mathcal{A}$ is the total surface area of the bounding walls. Thus, we have
\begin{equation}
f=f_1+f_2,    
\end{equation}
where 
\begin{equation}\label{}
\begin{aligned}
f_1 = \frac{1}{2\beta}
\int \frac{d^{2}\bold{q}}{(2\pi)^{2}}
\bigg(
\sum^{\infty}_{l=1}\ln\lambda'(\bold{q},0,l)
+\sum^{\infty}_{l=0}\ln\lambda(\bold{q},0,l)\bigg),
\end{aligned}
\end{equation}
\begin{equation}\label{}
\begin{aligned}
f_2 = \frac{2}{\beta}
\int \frac{d^{2}\bold{q}}{(2\pi)^{2}}
\sum^{\infty}_{l=0}{}'\sum^{\infty}_{n=1}\ln \lambda(\bold{q},n,l),
\end{aligned}
\end{equation}
\begin{equation}
\begin{aligned}
\lambda(\bold{q},n,l) = \textbf{q}^{2} + q_{l}^2 +\omega_n^2,\quad \lambda'(\bold{q},n,l) = \textbf{q}^{2} + q_{l}^2+ \omega_n^2+\varkappa^2,
\end{aligned}
\end{equation}
where $\varkappa = r_D^{-1}= (\beta\sum_a q_a^2n_a)^{1/2}$ is the inverse Debye length, $q_l=\pi l/L$ is the lateral photon momentum, and $\omega_n=2\pi n/\beta$ is the $n$th Matsubara frequency. A dashed sum means that the term with the number $l=0$ is included with a multiplier $1/2$.

\begin{figure}[h]
\centering
\includegraphics[width=0.7\textwidth]{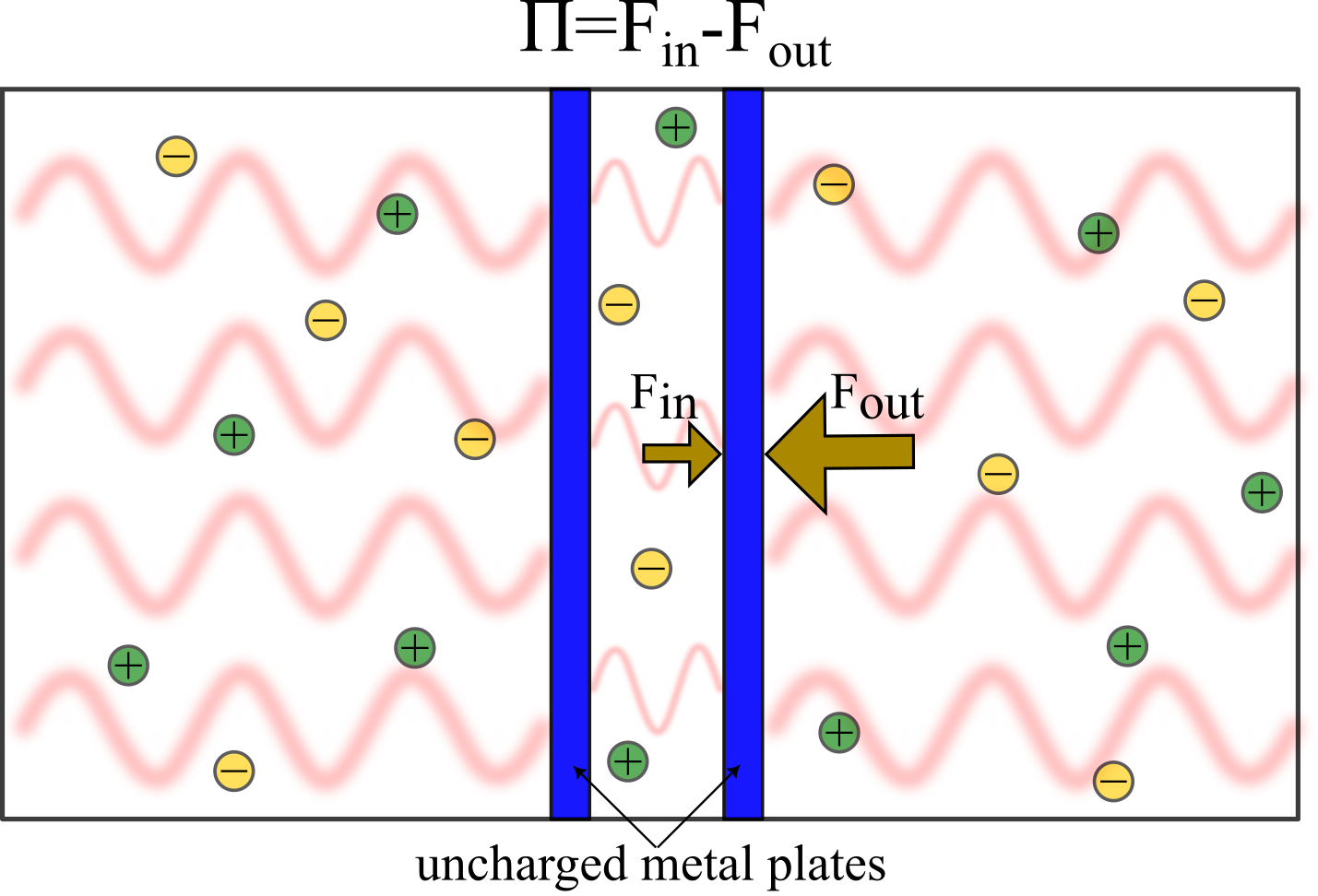}
\caption{A visual representation of the underlying mechanism of the Casimir effect.}
\label{Fig}
\end{figure}

Using the same method as in \cite{brandyshev2024finite}, we can sum over $l$ and differentiate with respect to $L$ to obtain the disjoining pressure -- {the excess pressure exerted on the wall of a pore compared to the pressure that would be exerted in the absence of walls, i.e., in a bulk system ~\cite{derjaguin1987surface,budkovstatistical},}
\begin{multline}
\label{Pi_1}
\Pi=-\left(\frac{\partial f}{\partial L}-\frac{\partial f}{\partial L}\bigg{|}_{L\to\infty}\right)\\  
=-\frac{T}{4\pi}\int\limits_{0}^{\infty} dq \;q \sqrt{q^2+\varkappa^2}\left(\coth\left(\sqrt{q^2+\varkappa^2}L\right)-1\right)-\frac{T}{4\pi}\int\limits_{0}^{\infty} dq \;q^2\left(\coth\left(qL\right)-1\right)\\-\frac{T}{\pi}\sum\limits_{n=1}^{\infty} \int\limits_{0}^{\infty} dq \;q\;\sqrt{q^2+\omega_n^2}\left(\coth\left(\sqrt{q^2+\omega_n^2}L\right)-1\right).
\end{multline}
{It is not difficult to understand that the disjoining pressure (\ref{Pi_1}) is equal to the total (Casimir) force acting on the relatively large parallel metallic plates immersed in a Coulomb gas, divided by their surface area (see Fig. \ref{Fig}).}

The first term on the right-hand side of equation (\ref{Pi_1}) represents the contribution from electric field fluctuations. Based on the Bohr-Van Leeuwen theorem, only the pure electric component of the disjoining pressure is influenced by the presence of classical charged particles in a volume between metal walls. That is why this contribution depends on the inverse Debye length, $\varkappa$. The second and third terms in the expression for the electromagnetic pressure arise from fluctuations that are unaffected by the presence of classical ions. {Such a contribution was also obtained by Maia Neto et al. \cite{maia2019scattering} for the case of an electrolyte solution confined between dielectric plates using the scattering formalism \cite{emig2007casimir}. Note also that the first and second zero Matsubara frequency terms on the right-hand side of equation (\ref{Pi_1}) can be derived from equation (25) in the paper by Maia Neto et al. \cite{maia2019scattering}, in the limit of metallic walls (i.e., an infinite dielectric constant), by differentiating the obtained free energy with respect to the inter-wall distance. Furthermore, the presence of the unscreened zero Matsubara frequency term was confirmed recently by experiments measuring the colloidal interaction between two silica microspheres in an aqueous solution at distances ranging from 0.2 to 0.5 microns using optical tweezers~\cite{pires2021probing}.}

Eq. (\ref{Pi_1}) can be rewritten as follows. Let us subtract and add the second term. This results in a purely repulsive ion contribution to the disjoining pressure, 
\begin{equation}
\Pi_{ion}=-\frac{T}{4\pi}\int\limits_{0}^{\infty} dq \;q \sqrt{q^2+\varkappa^2}\left(\coth\left(\sqrt{q^2+\varkappa^2}L\right)-1\right)+\frac{T}{4\pi}\int\limits_{0}^{\infty} dq \;q^2\left(\coth\left(qL\right)-1\right),
\end{equation}
which can be derived within the statistical field theory of the Coulomb gas~\cite{budkovstatistical,ninham1997ion,budkov2024thermomechanical} using image-charge concept, and the negative contribution of the electromagnetic field fluctuations, which coincides exactly with the result of Schwinger and coworkers~\cite{schwinger1978casimir}:
\begin{equation}
\label{EM}
\Pi_{EM}=-\frac{T}{\pi}\sum\limits_{n=0}^{\infty}{}^{'} \int\limits_{0}^{\infty} dq \;q\;\sqrt{q^2+\omega_n^2}\left(\coth\left(\sqrt{q^2+\omega_n^2}L\right)-1\right),
\end{equation}
so that
\begin{equation}
\Pi=\Pi_{ion}+\Pi_{EM}.
\end{equation}
Thus, our first-principle calculations justify the {\sl separation hypothesis} discussed in the Introduction. {It should be noted that recent studies within the framework of scattering formalism~\cite{maia2019scattering} have shown that ions in electrolyte solutions have a negligible effect on the electromagnetic component of the Casimir force. Our findings support this conclusion, as we have demonstrated that classical ions in Coulomb gas have no impact on the electromagnetic Casimir force. To fully understand the complex influence of ions on this phenomenon, we need to take into account quantum effects of ions, while not neglecting the terms associated with the $\boldsymbol{\xi}_a(\tau)$ and $\dot{\boldsymbol{\xi}}_a(\tau)$ in eq. (\ref{Z_tot}). This is beyond the scope of the current paper, so we will address it in separate context.}

Note that $\Pi_{ion}$ vanishes when $\varkappa=0$. Note also that it can be simplified to
\begin{equation}
\Pi_{ion}=\frac{T}{8\pi L^3}\left(\zeta(3)-\frac{1}{2}\int\limits_{2\varkappa L}^{\infty}  \;
\frac{dy \;y^2}{e^{y}-1}\right).
\end{equation}
The ion contribution to the disjoining pressure has the following asymptotic behavior
\begin{equation}
\label{ion_disj_press_asymp}
\Pi_{ion}\simeq
\begin{cases}
\frac{T\zeta(3)}{8\pi L^3}, & \varkappa L\gg 1\,\\
\frac{T\varkappa^2}{8\pi L},&\varkappa L\ll 1.
\end{cases}
\end{equation}

The electromagnetic contribution can be also simplified to~\cite{schwinger1978casimir}
\begin{equation}
\Pi_{EM}=- \frac{T}{4\pi L^3} \sum^{\infty}_{n=0}{}{}^{'} \;\int\limits_{2\omega_n L}^{\infty} \;
\frac{dy \;y^2}{e^{y}-1}.
\end{equation}

\section{Discussion}

Now let us discuss which formula correctly describes the Casimir disjoining pressure in a high-temperature regime in the absence of ions - the Schwinger or the Lifshitz formula. As discussed in the Introduction, when applied to the limit of ideally conductive walls, Lifshitz's theory encounters an ambiguity that requires taking the limits: $\omega_n\to 0$ and $\epsilon(\omega)\to \infty$. Schwinger et al. postulated the following order: $\epsilon(\omega) \to \infty$ and then $\omega_n\to 0$. At first glance, there seem to be no fundamental reasons why this particular order should be preferred over the one discussed in the works of Lifshitz and co-workers\cite{dzyaloshinskii1961general}. However, two main arguments support Schwinger's rule as the correct one.

First, from general physical considerations, in the limit of an ideal metal ($\epsilon(\omega)\to \infty$), the specific properties of the boundary surfaces become insignificant, and we obtain a pure electromagnetic field decoupled from matter in equilibrium with perfectly reflecting flat surfaces. This pure quantum field system, in its dynamics, must obey the symmetry of four-dimensional spacetime with respect to the Poincare group, $\text{ISO(3,1)}$. In thermodynamic equilibrium, the symmetry of four-dimensional spacetime reduces to that of four-dimensional Euclidean space with respect to the $\text{ISO(4)}$ group, where time is replaced by the imaginary time (inverse temperature). As we have shown in our previous work~\cite{brandyshev2024finite}, the Schwinger's formula follows from the Helmholtz free energy divided by temperature (or partition function)~\cite{brandyshev2024finite} which is invariant under the $\text{ISO(4)}$ group. At the same time, the expression obtained by Lifshitz et al. does not correspond to the partition function that is invariant under $\text{ISO(4)}$. Thus, we conclude that the disjoining pressure obtained by Schwinger respects the symmetries of four-dimensional space. The same can be said about the tangential pressure and, therefore, surface tension that we have recently obtained~\cite{brandyshev2024finite}. It is important to note that, due to the $\text{ISO(4)}$ symmetry of the partition function, we can obtain the low-temperature expansion of the disjoining pressure, as obtained by Mehra~\cite{mehra1967temperature}, from a resummation of Schwinger's high-temperature expansion~\cite{schwinger1978casimir,brandyshev2024finite}.

The second argument is based on the fundamental principles of classical electrodynamics and statistical mechanics. The difference between the results obtained by Schwinger et al. and those derived from Lifshitz's theory is due to the fact that the latter does not account for classical radiation pressure, which occurs in the high-temperature limit. In contrast, Schwinger's approach does take this into consideration. Indeed, the zero Matsubara frequency term in the total electromagnetic disjoining pressure~\cite{schwinger1978casimir, brandyshev2024finite}, 
\begin{equation}
\label{Pi_cl}
\Pi_{EM}^{(n=0)}=-\frac{T}{2\pi}\int\limits_{0}^{\infty} dq \;q^2\left(\coth\left(qL\right)-1\right)=-\frac{T\zeta (3)}{4\pi L^3},
\end{equation}
is simply the disjoining pressure generated by equilibrium classical radiation. This is demonstrated using classical electrodynamics and statistical mechanics in Appendix C.

{It is important to note that recent highly accurate experiments~\cite{bimonte2021measurement} have been conducted to measure the Casimir force between metal plates. These experiments showed that the plasma model for metal, which allows neglecting the imaginary part of the dielectric constant~\cite{klimchitskaya2023casimir,jackson2021classical}, is accurate for short separations up to 4.8 microns. This assumption is consistent with the evaluation of the electromagnetic part of the disjoining pressure (\ref{EM}) obtained for the ideal metal case. However, for separations greater than 4.8 microns, definitive conclusions could not be draw from the experimental data due to the limitations of the measurement techniques used. Therefore, further experimental confirmation is needed to verify the theory at larger distances.}

{We would like to share some reflections on the current results and their potential future developments. As noted in the Introduction, approaches based on a purely electrostatic consideration \cite{jancovici2005casimir,jancovici2004screening,buenzli2005microscopic,budkov2024thermomechanical} yield absolute values for the zero Matsubara mode of the Casimir force that are half of those predicted by quantum electrodynamics in the high-temperature approximation for an ideal metal walls. This is not surprising from a physics standpoint, as the electrostatic approach clearly cannot provide a comprehensive description of the phenomenon. This is because it ignores the effects related to the equilibrium electromagnetic field. In fact, as we have shown above (see also ref. \cite{kaufman2024lectures}), in the classical approximation, the total partition function of a system of charges interacting with an electromagnetic field can be factorized into the configuration partition function of charges interacting via the Coulomb potential and the partition function of the electromagnetic field itself. Therefore, these two partition functions must be calculated separately. Note also that we have examined only the behavior of the ions in a vacuum between ideal metal walls. However, it is important to note that this represents an idealization, as in real-world scenarios, ions are typically surrounded by polar solvent molecules in electrolyte solutions~\cite{maia2019scattering,seyedzahedi2021effects}. This simplification allowed us to illustrate, based on the first principles, how the presence of charged particles affects the Casimir force at high temperatures. To incorporate the effects of a solvent's dielectric medium, we can introduce a phenomenological dielectric function, in line with standard Lifshitz theory~\cite{podgornik1989forces,podgornik1988solvent,podgornik2003reformulation,netz2001static,klimchitskaya2023casimir} or scattering formalism~\cite{maia2019scattering,emig2007casimir,seyedzahedi2021effects}. To accomplish this, it is necessary to develop a finite-temperature quantum field theory that includes an action for the electromagnetic field in a dielectric medium with a specific phenomenological dielectric function. These issues deserve separate consideration in future publications, within the framework of finite-temperature quantum field theory.}

\section{Conclusions}

In conclusion, our justification for the separation hypothesis is grounded in the fundamental principles of quantum electrodynamics and quantum statistical physics. We have shown that, in the Gaussian approximation, the positive ionic contribution to the Casimir force, of electrostatic origin, can be separated from the electromagnetic fluctuations responsible for the original Casimir effect. It is crucial to highlight the significance of calculating the zero Matsubara mode term in the electromagnetic contribution using the formula put forth by Schwinger et al., rather than that of Lifshitz et al. This choice is supported by the $\text{ISO(4)}$ symmetry of the Helmholtz free energy, from which Schwinger's formula can be derived, and by the correspondence between the zero Matsubara mode in Schwinger's expansion and the pressure of classical electromagnetic radiation in equilibrium with conductive walls, as explained by classical electrodynamics and statistical physics.

\appendix
\section{Gauge invariance of partition function}
The partition function of the described system can be written as follows
\begin{equation}
\label{InvariantZ}
\mathcal{Z}=\oint \mathcal{D}A\oint d\sigma[\{\bold{r}_{i}^{(+)}\}]\oint d\sigma[\{\bold{r}_{j}^{(-)}\}] e^{\mathcal{S}},
\end{equation}
where
\begin{equation}
\mathcal{S}=\int\limits_{0}^{\beta}d\tau\int d^3\mathbf{x} \; \bigg[\mathscr{L}_{EM}-i(\hat{\rho}\varphi-\hat{\bold{j}}\cdot\bold{A})
\bigg],
\end{equation}
is the total action with the Lagrangian of the electromagnetic field
\begin{equation}\label{}
\mathscr{L}_{EM} = -\frac{1}{4} F_{\mu\nu}F_{\mu\nu},
\end{equation}
where $F_{\mu\nu}=\partial_{\mu}A_{\nu} - \partial_{\nu}A_{\mu}$ is the electromagnetic tensor.
Using the electric charge conservation law,
\begin{equation}
\partial_{\tau}\hat{\rho}+\nabla\cdot \hat{\bold{j}}=0,    
\end{equation}
we can easily check that (\ref{InvariantZ}) is a gauge invariant.

To fix gauge, let us consider the identity
\begin{equation}\label{oneIdentity}
\begin{aligned}
\oint\mathcal{D}\theta \, |D| \, \delta [\partial_{\mu}A^{\theta}_{\mu}(x) - B(x)] = 1,
\end{aligned}
\end{equation}
where $|D|$ is determined by (\ref{DetFaddeev}), $A^{\theta}_{\mu}$ is defined by (\ref{Atransformed}), and $B(x)$ is an arbitrary function; $\delta[\cdot]$ is the functional generalization of Dirac's delta-function. Using the identity (\ref{oneIdentity}), the partition function can be rewritten as
\begin{equation}
\label{Z2}
\mathcal{Z}=\oint\mathcal{D}\theta\oint \mathcal{D}A\oint d\sigma[\{\bold{r}_{i}^{(+)}\}]\oint d\sigma[\{\bold{r}_{j}^{(-)}\}] |D| \, \delta [\partial_{\mu}A^{\theta}_{\mu} - B]e^{\mathcal{S}}.
\end{equation}
Calculation of the functional integral over all possible functions $B(x)$ and dividing by gauge group volume $\mathcal{V}_{\theta} = \oint \mathcal{D}\theta$,
\begin{equation}\label{}
\begin{aligned}
Z =\mathcal{V}_{\theta}^{-1} \oint\mathcal{D}B\; \mathcal{Z}\exp \bigg(-\frac{\alpha B^{2}}{2}\bigg),
\end{aligned}
\end{equation}
results in (\ref{Z}). From (\ref{oneIdentity}) it follows that $\mathcal{Z}$ in (\ref{Z2}) does not depend on $B(x)$ thus the integration over $B(x)$ is equivalent to multiplying by infinite constant~\cite{brandyshev2024finite}.

\section{Calculation of partition function}
The partition function at $\alpha=1$ can be written as
\begin{equation}
\quad Z= |D|(\Delta_{0}\Delta_{1}\Delta_{2}\Delta_{3})^{-\frac{1}{2}}
\end{equation}
where $\Delta_{\mu}$ is the determinant corresponding to the functional integrations over $A_{\mu}$. Let us consider the eigenfunctions of the D'Alembert operator 
\begin{equation}\label{}
\begin{aligned}
-\square \; v(x,\bold{q},n,l) = \lambda(\bold{q},n,l) v(x,\bold{q},n,l),\\
l = 1,2\ldots ,\quad n = 0,\pm 1,\pm 2, \ldots 
\end{aligned}
\end{equation}
satisfying the boundary conditions at $x_3=0$ and $x_3=L$
\begin{equation}
v=0,    
\end{equation}
with eigenvalues
\begin{equation}\label{}
\begin{aligned}
\lambda(\bold{q},n,l) = \textbf{q}^{2} + q_{l}^2 +\omega_n^2,
\end{aligned}
\end{equation}
The determinant $|D|$ can be rewritten as
\begin{equation}\label{DetFaddeev}
|D| = \text{Det}\bigg(-\frac{\delta \square \theta}{\delta \theta}\bigg),
\end{equation}
where $\theta(x)$ is the parameter of the gauge transformation~\cite{kapusta2007finite,brandyshev2024finite}
\begin{equation}\label{Atransformed}
A^{\theta}_{\mu} = A_{\mu} - \partial_{\mu}\theta.
\end{equation}
It has been shown recently~\cite{brandyshev2024finite} that $\theta(x)$ is satisfied by the same boundary conditions as $v$, so that
\begin{equation}\label{}
|D|=\exp\bigg( \sum^{\infty}_{l=1} \ln\bar{\lambda}_{l}\bigg),
\end{equation}
where
\begin{equation}\label{}
\ln\bar{\lambda}_{l} = \mathcal{A}\int \frac{d^{2}\bold{q}}{(2\pi)^{2}} \sum_{n\in Z} \ln \lambda(\bold{q},n,l)
\end{equation}
and also $A_0$, $A_1$ and $A_2$ are satisfied by the same boundary conditions. Therefore
\begin{equation}\label{}
\begin{aligned}
\int \mathcal{D}A_{1} \, \exp\bigg(\frac{1}{2}\bigg(A_{1},\square A_{1}\bigg)\bigg)=\Delta_{1}^{-\frac{1}{2}},
\end{aligned}
\end{equation}
\begin{equation}
\Delta_{1}=|D|,    
\end{equation}
and for $A_2$ one gets the same result as for $A_1$, i.e. $\Delta_{1}=\Delta_{2}$.
Let us note that $v(x,\bold{q},0,l)$ are eigenfunctions of other operator
\begin{equation}\label{}
(\varkappa^2 - \square) \; v(x,\bold{q},0,l) = \lambda'(\bold{q},0,l) v(x,\bold{q},0,l),
\end{equation}
with other eigenvalues
\begin{equation}\label{}
\begin{aligned}
\lambda'(\bold{q},0,l) = \textbf{q}^{2} + q_{l}^2 +\varkappa^2,
\end{aligned}
\end{equation}
where
\begin{equation}
\label{}
\varkappa^2=\beta\sum\limits_{a=\pm}n_a q_a^2.
\end{equation}
From this, one can get
\begin{equation}\label{A14}
\begin{aligned}
\int \mathcal{D}\varphi \, \exp\bigg(\frac{1}{2}\bigg(\varphi,\square \varphi\bigg) - \frac{1}{2}\sum\limits_{a=\pm}q_a^2n_{a}\int d^3\bold{x}\bigg(\int\limits_{0}^{\beta}d\tau \varphi(\tau,\bold{x})\bigg)^2\bigg)=\Delta_{0}^{-\frac{1}{2}},
\end{aligned}
\end{equation}
where
\begin{equation}\label{expand1}
\begin{aligned}
\varphi(x)=\varphi(\tau,\bold{x})=\sum^{\infty}_{l=1}\sum_{n\in Z} \int d^{2}\bold{q} \; a_{0}(\bold{q},n,l)v(x,\bold{q},n,l) + h.c.,
\end{aligned}
\end{equation}
and after integrating $\varphi(x)$ with respect to $\tau$ in the expression (\ref{A14}) only one non-zero term with $n=0$ remains in sum (\ref{expand1}) because the integral of the function $v(x,\bold{q},n,l) \sim e^{i\omega_{n}\tau}$ is equal to zero at $n\neq 0$. Thus, we have
\begin{equation}\label{}
\begin{aligned}
\ln\Delta_{0} = \mathcal{A}\sum^{\infty}_{l=1}
\int \frac{d^{2}\bold{q}}{(2\pi)^{2}}
\bigg(\ln\lambda'(\bold{q},0,l)
+2\sum^{\infty}_{n=1}\ln \lambda(\bold{q},n,l)\bigg),
\end{aligned}
\end{equation}
$A_3$ satisfies other boundary conditions at $x_3=0$ and $x_3=L$ 
\begin{equation}
\partial_3A_3=0,    
\end{equation}
which gives the zero charge density on the surfaces of plates, i.e. $\nabla \cdot \bold{E} = 0$. Thus, let us consider the eigenfunctions 
\begin{equation}\label{}
\begin{aligned}
-\square \; u(x,\bold{q},n,l) = \lambda(\bold{q},n,l) u(x,\bold{q},n,l),\\
l = 0,1,2\ldots ,\quad n = 0,\pm 1,\pm 2, \ldots 
\end{aligned}
\end{equation}
with the boundary conditions at $x_3=0$ and $x_3=L$ 
\begin{equation}
\partial_3 u =0,   
\end{equation}
then one can get
\begin{equation}\label{}
\begin{aligned}
\int \mathcal{D}A_{3} \, \exp\bigg(\frac{1}{2}\bigg(A_{3},\square A_{3}\bigg)\bigg)=\Delta_{3}^{-\frac{1}{2}},
\end{aligned}
\end{equation}
where
\begin{equation}\label{}
\Delta_{3}=\exp\bigg( \sum^{\infty}_{l=0} \ln\bar{\lambda}_{l}\bigg).
\end{equation}
that can be rewritten as 
\begin{equation}\label{}
\begin{aligned}
\ln\Delta_{3} = \mathcal{A}\sum^{\infty}_{l=0}
\int \frac{d^{2}\bold{q}}{(2\pi)^{2}}
\bigg(\ln\lambda(\bold{q},0,l)+2\sum^{\infty}_{n=1}\ln \lambda(\bold{q},n,l)\bigg) 
\end{aligned}
\end{equation}

\section{Disjoining pressure of electromagnetic radiation in classical physics}
Let us condider the standing electromagnetic waves between perfectly conducting metal flat walls separated by distance $L$. We consider the Coulomb gauge, i.e. that $\varphi=0$ and $\nabla\cdot \bold{A}=0$. 
The general solution of the d'Alembert equation,
\begin{equation}
\frac{\partial^2\bold{A}}{\partial t^2}-\Delta \bold{A}=0,
\end{equation}
with boundary conditions on the walls corresponding to the ideal metal
\begin{equation}
\label{bound_1}
\partial_3 A_{3}(x_1,x_2,0,t)=\partial_3 A_{3}(x_1,x_2,L,t)=0, 
\end{equation}
\begin{equation}
\label{bound_2}
\bold{A}^{||}(x_1,x_2,0,t)=\bold{A}^{||}(x_1,x_2,L,t)=0,~\bold{A}^{||}=(A_1,A_2,0),
\end{equation}
satisfying the gauge relation $\nabla \cdot \bold{A}=0$, is determined by the well-known formula~\cite{barton1970quantum}
\begin{equation}
\label{A_solutioin}
\bold{A}=\sum\limits_{\bold{q}}\sum\limits_{l=0}^{\infty}{}^{''}\left(a_{\bold{q}l}^{(1)}[\bold{e}_{\bold{q}},\bold{e}_{3}]\sin(q_{l}x_3)+a_{\bold{q}l}^{(2)}\left(\bold{e}_{\bold{q}}\frac{iq_l}{\omega_{\bold{q}l}}\sin(q_{l}x_3)-\bold{e}_3\frac{|\bold{q}|}{\omega_{\bold{q}l}}\cos(q_l x_3)\right)\right)e^{i\bold{q}\cdot\boldsymbol{\rho}}+c.c.,
\end{equation}
where $\boldsymbol{\rho}=(x_1,x_2,0)$, $q_{l}=\pi l/L$, $l=0,1,2,\dots$, $\bold{q}=(2\pi n_x/L_x, 2\pi n_y/L_y,0)$, $n_{x},n_{y}\in Z$, $\omega_{\bold{q}l}=\sqrt{\bold{q}^2+q_l^2}$, $\bold{e}_{\bold{q}}=\bold{q}/|\bold{q}|$, $\bold{e_3}$ is the unit vector along the $x_3$-axis, $[\bold{e}_{\bold{q}},\bold{e}_{3}]$ is the vector product; the double prime on the sum means that there is an extra factor of $1/\sqrt{2}$ for the term with $l = 0$.

Note that the coefficients $a_{\bold{q}l}^{(\alpha)}$ (the index $\alpha = 1, 2$ indicates the polarization direction) are the functions of time and satisfy the equations
\begin{equation}
\ddot{a}_{\bold{q}l}^{(\alpha)}+\omega_{\bold{q}l}^2a_{\bold{q}l}^{(\alpha)}=0,
\end{equation}
i.e. $a_{\bold{q}l}^{(\alpha)}\sim e^{-i\omega_{\bold{q}l}t}$.

The electric and magnetic fields are determined by relations
\begin{equation}
\label{fields}
\bold{E}=-\dot{\bold{A}},~\bold{B}=[\nabla,\bold{A}],
\end{equation}
which yield
\begin{equation}
\label{el_f}
\bold{E}=\sum\limits_{\bold{q}}\sum\limits_{l=0}^{\infty}{}^{''}\left(-a_{\bold{q}l}^{(1)}i\omega_{\bold{q}l}[\bold{e}_{\bold{q}},\bold{e}_{3}]\sin(q_{l}x_3)+a_{\bold{q}l}^{(2)}\left(\bold{e}_{\bold{q}}{q_l}\sin(q_{l}x_3)+i\bold{e}_3|\bold{q}|\cos(q_l x_3)\right)\right)e^{i\bold{q}\cdot\boldsymbol{\rho}}+c.c.,
\end{equation}
\begin{equation}
\label{mag_f}
\bold{B}=\sum\limits_{\bold{q}}\sum\limits_{l=0}^{\infty}{}^{''}\left(-a_{\bold{q}l}^{(2)}i\omega_{\bold{q}l}[\bold{e}_{\bold{q}},\bold{e}_{3}]\cos(q_{l}x_3)+a_{\bold{q}l}^{(1)}\left(\bold{e}_{\bold{q}}{q_l}\cos(q_{l}x_3)-i\bold{e}_3|\bold{q}|\sin(q_l x_3)\right)\right)e^{i\bold{q}\cdot\boldsymbol{\rho}}+c.c.
\end{equation}

The total energy of the electromagnetic field is
\begin{equation}
H=\frac{1}{2}\int d^{3}\bold{r}\left(\bold{E}^2+\bold{B}^2\right).
\end{equation}
Using eqs. (\ref{A_solutioin}), (\ref{el_f}) and (\ref{mag_f}), integration yields
\begin{equation}
H=\frac{V}{2}\sum\limits_{\alpha=1,2}\sum\limits_{\bold{q}}\sum\limits_{l=0}^{\infty}{}^{'}(q_{l}^2+\bold{q}^2)a_{\bold{q}l}^{(\alpha)}{}^{*}a_{\bold{q}l}^{(\alpha)}
\end{equation}
where $V=L_x L_yL$ is the total volume enclosed by the metal walls.

Then, introducing the canonical variables
\begin{equation}
{Q}_{\bold{q}l}^{(\alpha)}=\frac{\sqrt{V}}{2}({a}_{\bold{q}l}^{(\alpha)}+{a}_{\bold{q}l}^{(\alpha)}{}^{*}),
\end{equation}
\begin{equation}
{P}_{\bold{q}l}^{(\alpha)}=-i\omega_{\bold{q}l}\frac{\sqrt{V}}{2}({a}_{\bold{q}l}^{(\alpha)}-{a}_{\bold{q}l}^{(\alpha)}{}^{*}),
\end{equation}
where the symbol ${}^{*}$ means the complex conjugation, we can rewrite the total energy of the electromagnetic field in their terms as follows
\begin{equation}
H[P,Q]=\frac{1}{2}\sum\limits_{\alpha=1,2}\sum\limits_{\bold{q}}\sum\limits_{l=0}^{\infty}{}^{'}\left({{P}_{\bold{q}l}^{(\alpha)}}^2+\omega_{\bold{q}l}^2{{Q}_{\bold{q}l}^{(\alpha)}}^2\right).
\end{equation}
Note that these variables are canonical because they satisfy the Hamilton's equations:
\begin{equation}
\dot{Q}_{\bold{q}l}^{(\alpha)}=\frac{\partial H}{\partial{P}_{\bold{q}l}^{(\alpha)}},~\dot{P}_{\bold{q}l}^{(\alpha)}=-\frac{\partial H}{\partial{Q}_{\bold{q}l}^{(\alpha)}}.
\end{equation}

For the harmonic oscillators of the electromagnetic field, which are in equilibrium with metal walls, we have the following Gibbs distribution function
\begin{equation}
\mathcal{P}\left[P,Q\right]=\frac{1}{Z}e^{-\beta H\left[P,Q\right]}.
\end{equation}
The Helmholtz free energy of the electromagnetic field is
\begin{equation}
F=-T\ln Z,
\end{equation}
where the partition function is
\begin{equation}
Z=\int \mathcal{D}P\, \mathcal{D}Q\, e^{-\beta H\left[P,Q\right]}.
\end{equation}
Calculating the Gaussian integrals, we obtain the Helmholtz free energy up to infinite constant which does not depend on $L$:
\begin{equation}
F=T\sum\limits_{\bold{q}}\sum\limits_{l=0}^{\infty}\ln (\bold{q}^2+q_l^2).
\end{equation}
The electromagnetic radiation pressure is 
\begin{equation}
P=-\frac{\partial (F/\mathcal{A})}{\partial L},
\end{equation}
where $\mathcal{A}=L_x L_y$ is the total area of the metal walls. Therefore, we obtain
\begin{equation}
P=-\frac{2T}{L}\int\frac{d^2\bold{q}}{(2\pi)^2}\sum\limits_{l=1}^{\infty}\frac{q_l^2}{\bold{q}^2+q_l^2}.
\end{equation}
where we have used the fact that for very large walls we have
\begin{equation}
\sum_{\bold{q}}(\cdot)\to \mathcal{A}\int \frac{d^2\bold{q}}{(2\pi)^2}(\cdot).
\end{equation}

Using the same approach to regularizing divergent sums as in \cite{brandyshev2024finite}, we obtain
\begin{equation}
P=-T\int\frac{d^2\bold{q}}{(2\pi)^2}q\coth (qL).
\end{equation}
In agreement with (\ref{Pi_cl}), the radiation disjoining pressure is
\begin{equation}
\Pi_{EM}^{(cl)} = P-P|_{L\to \infty}=\Pi_{EM}^{(n=0)}=-T\int\frac{d^2\bold{q}}{(2\pi)^2}q \left(\coth (qL)-1\right)=-\frac{T\zeta(3)}{4\pi L^3}.
\end{equation}

\section*{Acknowledgments}
The authors thank the Russian Science Foundation (Grant No. 24-11-00096) for financial support.

\selectlanguage{english}
\bibliography{name}
\end{document}